\documentclass[twocolumn,amsmath,showpacs]{revtex4}
\usepackage{graphicx}
\usepackage{xcolor}

\begin{document}
\title{
Synchronization of Synchrotron Radiation Bursts during a spatio-temporal Instability in accelerator-Based source}

\author{
  C. Evain$^{1}$, A.-A. Diallo$^{1,2}$, E. Roussel$^{1}$, C. Szwaj$^{1}$,  M.~Herda$^{2}$,   M.-A. Tordeux$^{3}$, F. Ribeiro$^{3}$, M. Labat$^{3}$, N. Hubert$^{3}$, J.-B. Brubach$^{3}$, P. Roy$^{3}$,  S. Bielawski$^{1}$}

\affiliation{1. Univ. Lille, CNRS, UMR 8523 - PhLAM - Physique des Lasers Atomes et Molécules,  F-59000 Lille, France}
\affiliation{2: Inria, Univ. Lille, CNRS, UMR 8524 - Laboratoire Paul Painlevé, F-59000 Lille, France}
\affiliation{3: Synchrotron SOLEIL, Saint Aubin, BP 34, 91192 Gif-sur-Yvette, France}

\begin{abstract}

  Synchronization is a fundamental phenomenon in dynamical systems, occurring in a wide range of contexts such as mechanical, chemical, biological, and social systems.
  In this work, we explore a novel manifestation of synchronization in accelerator-based light sources, specifically in storage rings where relativistic electron bunches circulate and emit synchrotron radiation, used for user experiments.
  In such systems, a systematic spatio-temporal instability arises when the bunch contains a large number of electrons.
  This instability is characterized by the spontaneous formation of microstructures within the bunch, which appear with a bursting behavior.
  We demonstrate that these bursting events can be synchronized with an external sinusoidal signal by modulating the electric field in a radiofrequency (RF) cavity.
  This external modulation induces typical synchronization features such as Arnold tongues at fundamental, harmonic, and subharmonic frequencies of the natural bursting rate, as well as phase-slip phenomena near the synchronization threshold.
  The synchronization mechanism is analyzed using numerical simulations based on the Vlasov–Fokker–Planck equation, and a proof-of-principle experiment is conducted at the SOLEIL synchrotron facility.
\end{abstract}
 
\date{\today}
\maketitle

\section{Introduction}
Synchronization is a general phenomenon that arises when systems exhibiting periodic or quasi-periodic behavior adjust their rhythms due to mutual interaction or external forcing~\cite{Pikovsky2001,Balanov2009}.
Two main scenarios are typically considered in the study of synchronization.
In the first, several coupled systems, each with an intrinsic oscillatory behavior, synchronize with each other, even if the coupling is extremely weak — as famously demonstrated by Huygens in the case of pendulum clocks~\cite{Pikovsky2003}.
In the second scenario, a single system, with a periodic  or quasi-periodic by nature, synchronizes with an external periodic force.
This externally driven synchronization is  characterized by the emergence of Arnold tongues, which delimit the regions of effective synchronization in the parameter space defined by the modulation frequency and amplitude (for example:~\cite{Arnold1961, Glass1983, Pikovsky1997, Ruhunusiri2012, Golden2021}).\\

In this work, we investigate a synchronization process that occurs in an accelerator-based light source, using an external forcing.
The accelerator is a storage rings, where relativistic electron bunches circulate and emit synchrotron radiation [see FIG.~1(a)].\\

Theses facilities are among the most commonly employed accelerator-based light sources.
In such systems, electron bunches travel at relativistic speeds along a quasi-periodic orbit, with circumferences ranging from several tens to hundreds of meters.
When the electrons are deflected from their straight-line trajectories by magnetic fields, such as dipole bending magnets or undulators, they emit synchrotron radiation.
This radiation is extracted and delivered to users for scientific experiments.\\

In these systems, a spatio-temporal instability systematicaly emerges when a bunch contains a sufficiently large number of electrons.
A portion of the emitted synchrotron radiation interacts with other electrons within the bunch, particularly in bending magnets, as illustrated in figure~1(b).
Once the electron number in a bunch exceeds a thershold value, this interaction leads to the spontaneous formation of structures in the longitudinal distribution of the bunch.
This is shown in Fig.~1(g), which displays the longitudinal phase space (i.e. electrons distribution versus longitudinal position and energy) where these structures can be observed.\\

These microstructures enable the electrons in the bunch to emit synchrotron radiation coherently, i.e., in phase, at frequencies corresponding to the wavelengths of the modulations in the longitudinal charge density. This coherent synchrotron radiation (CSR) is typically emitted in the terahertz (THz) range and can reach power levels several orders of magnitude higher than standard incoherent synchrotron radiation, which is observed in the absence of such microstructures.\\

\begin{figure*}[htpb!]
  \includegraphics[width=1\linewidth]{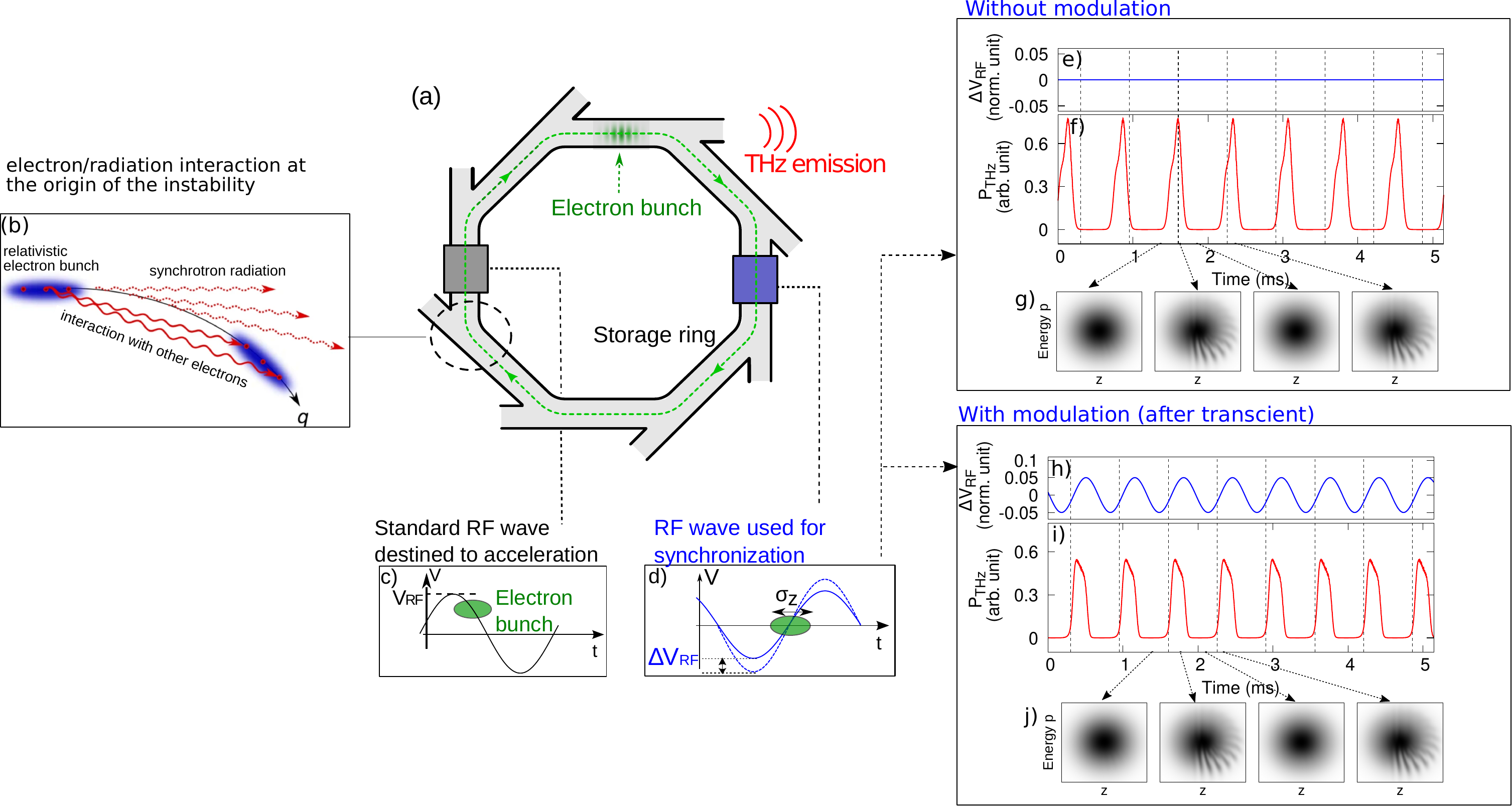}
  \caption{
Schematic representation of the system and of the synchronization phenomena (based on numerical simulations of the Vlasov-Fokker-Planck equation, see Appendix for details).  
(a) An electron bunch circulates in a storage ring at relativistic velocity and emits synchrotron radiation.  
(b) A portion of this radiation interacts with other electrons in the bunch (in particular in bending magnets), leading to the microbunching instability when the number of electrons exceeds a threshold.  
This spatio-temporal instability is characterized by the spontaneous appearance of structures in the longitudinal distribution of the bunch [see electron bunch phase-space in (g) and (j)]. (f) These structures appear by bursts, resulting in intense bursts of coherent synchrotron radiation (CSR). Note that the burst period $T_b$ is much larger that the revolution period of the bunch in the ring $T_r$ (for example at SOLEIL, $T_b\simeq~2$~ms and $T_s\simeq~1~\mu$s). 
d) To synchronize theses bursts, an external modulation signal $\Delta V_{RF}$ is applied to a RF cavity, to modulate the amplitude of the RF signal. It  permits to modify the longitudinal dimensions of the bunch (which are directly related to the instability gain). i) example of the emitted CSR when the modulation shown in h) is applied. j) associated distribution of the electron bunch in the longitudinal phase-space.} \label{fig:setup} 
\end{figure*}

One important aspect of this instability is that the structures - and thus the coherent emission -  don't appear continuously but in the form of bursts [see FIG.~1(f),(g)].
As explained in more details after, this phenomena comes from the variation of the bunch length during the instability.
In this article, we  show that these bursts can be synchronized on an external signal, which modulate the electric field in a RF cavity.\\

The RF cavities are used to accelerate the electrons in order to compensate for the energy loss due to radiation [see Fig.~1(c)]. However, they also affect the bunch length.
By varying the RF field inside these cavities [see Fig.~1(d)], it is then possible to manipulate the bunch dynamics during the micro-bunching instability.
It has been shown that this type of method can be used to suppress the bursting behavior while preserving the micro-structures (when combined with a feedback loop)~\cite{Evain2019, Evain2023}; and also to obtain much more intense bursts, compared to the spontaneous bursts (a process called gain-switching, based on turning the modulation ON and OFF)~\cite{Evain2024}.
In this work, we show that a synchronization of the bursts can be achieved through a sinusoidal modulation of the RF signal amplitude [see Fig.~1(h),(i)].\\

Note also that bursting can be triggered not only through RF modulation but also via interactions between the electron beam and an external laser, although no synchronization phenomenon has been observed in this case~\cite{Byrd2006, Roussel2014b}.\\

In this article, the first part is dedicated to describe the fundamental mechanisms behind the observed phenomena, supported by numerical simulations; and the second part presents experimental results at the Synchrotron SOLEIL.

%%%%%% Principle %%%%%%%%%%%%%

\section{Synchronization principle}

\begin{figure*}[htpb!]
  \includegraphics[width=1\linewidth]{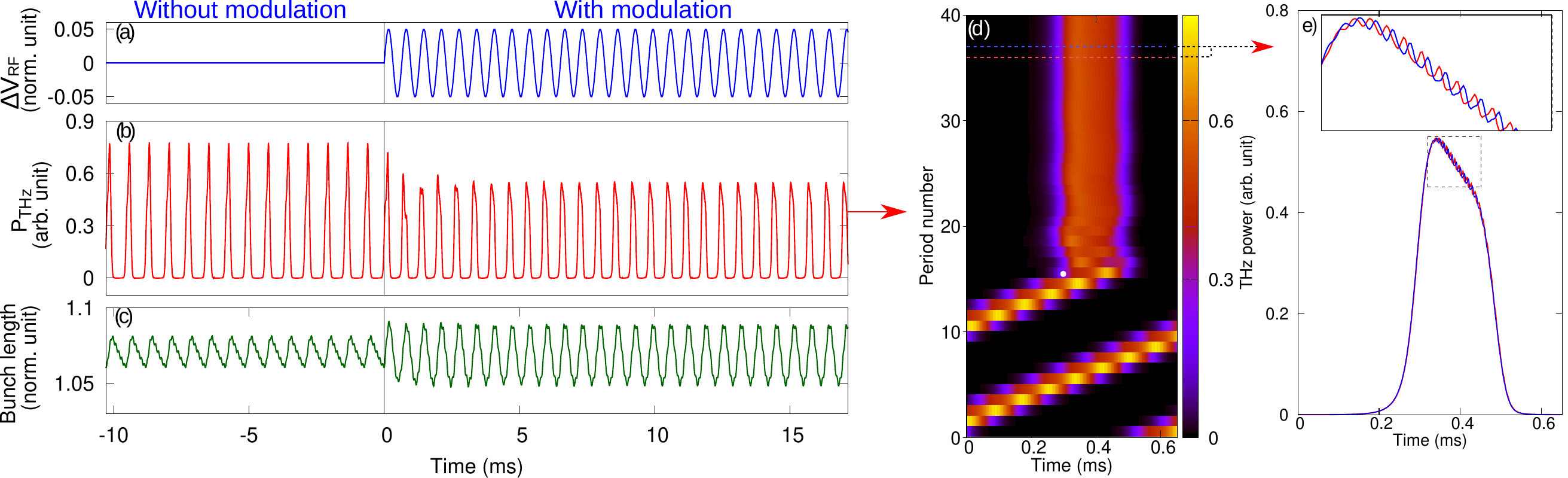}
  \caption{Illustration of the synchronization phenomena from numerical simulations (from the integration of the Vlasov-Fokker-Planck equation). a-c) temporal evolution of the modulation of the RF signal amplitude (a), the emitted THz power (b) and the bunch length (c), when the RF modulation is turn ON at $t=0$, with $T_{RF}\simeq 0.65~$ms  and $A_{RF}=0.05$ (the spontaneous burst period is about 0.73~ms). d) THz power (same data than in b) in a map representation, where each horizontal line corresponds to a period of the RF signal modulation (the white dot indicates the time at which the modulation is applied). e-f) Details of two bursts, where a modulation  can be observe, which is due to the evolution of the micro-structures in the phase-space~\cite{Roussel2014}. 
  } \label{fig:dyn_no_sync}
  %278.3/13/2/pi/4640*1000=0.73
  %0.789*19/2/pi/4.64=0.5142  (0.789=300/19 - 15)
  %0.789*19 + 6.28324 - 0.21562*2*pi*4.64= 14.988  (6.28324: initial value in the map) - 0.21562: line removed in the map

\end{figure*}

The description of the instability and the associated synchronization process is based on the numerical integration of the Vlasov–Fokker–Planck equation.
This equation is known to capture important features of the dynamics of an electron bunch during the microbunching instability~\cite{Venturini2002, Venturini2005, Roussel2014}.
The model includes the natural evolution of the bunch in the longitudinal phase space and the interaction of the electrons with their radiation (at the origin of the instability) through a simplified model (the so-called parallel plate model). Details on the model are given in appendix and in~\cite{Roussel2014}.\\

During the spatio-temporal instability, structures appears spontaneously in the longitudinal phase space of the bunch, as shown in Fig.~1(g).
These structures lead to the emission of a strong coherent synchrotron radiation at frequencies determined by the characteristic wavelength of the structures (here in the THz domain).
Since these structures are usually present by bursts, the coherent emission is also emitted by burts, as shown in FIG.~2(b) (for $t<0$, when the RF modulation is off).
This bursting behavior is due to a variation of the bunch length during the instability.
As shown in Fig.~2(h), the apparition of microstructures causes the emission of CSR, but it also induces an elongation of the bunch.
As the bunch length increases, the peak charge density decreases, causing the system to drop below the instability threshold.
In this regime, the microstructures doesn't appear anymore, and no CSR is emitted anymore.
Then, due to natural radiation damping — from incoherent synchrotron radiation — the bunch length gradually shortens [see FIG.~2(c)].
When the bunch becomes sufficiently short, the system once again crosses the instability threshold, allowing new microstructures to form and initiating another burst of CSR.
This cyclical process gives rise to the bursting behavior of the coherent THz emission.\\

This description of the electron bunch dynamics concerns the undisturbed system, i.e. without any modulation.
To obtain the synchronization process, the signal in a RF cavity is changed.
More precisely, it is the slope of the RF signal which is modified, since - as explained above -  the bunch length evolution is responsible of the bursting behavior we want to synchronize, and this bunch length is directly linked to the RF signal slope (a higher slope tends to contract the bunch and inversely). 
In this study, the RF signal slope is modulated with a sinusoidal signal $\Delta V_{RF}$ [see FIG.~2(a)], with an amplitude $A_{RF}$ and a period $T_{RF}$ : $\Delta V_{RF}= A_{RF} \sin\left(\frac{2\pi}{T_{RF}} t\right) $.\\

With this configuration, it is possible to observe that when the modulation is applied - and after a short transient -  a synchronization phenomena appears : the bursts are synchronized with the modulation signal [see FIG.~2(a),(b)].
This can be clearly observed with a map representation [see FIG.~2(d)], where the evolution of the THz power is cut in segments of period of $T_{RF}$; and where the different segments are supperposed vertically (with the time during a period in the horizontal axe and the amplitude of the THz power in color-scale).
In this representation -  and after the transient -  we observe that the burst appears in phase with the modulation signal, which is not the case when the modulation amplitude is zero (causing a drift of the burst positions in this representation).
We observe also that the bunch length variation is  locked on the modulation signal [cf. FIG~2(c)].\\

\begin{figure*}[htpb!]
  \includegraphics[width=0.95\linewidth]{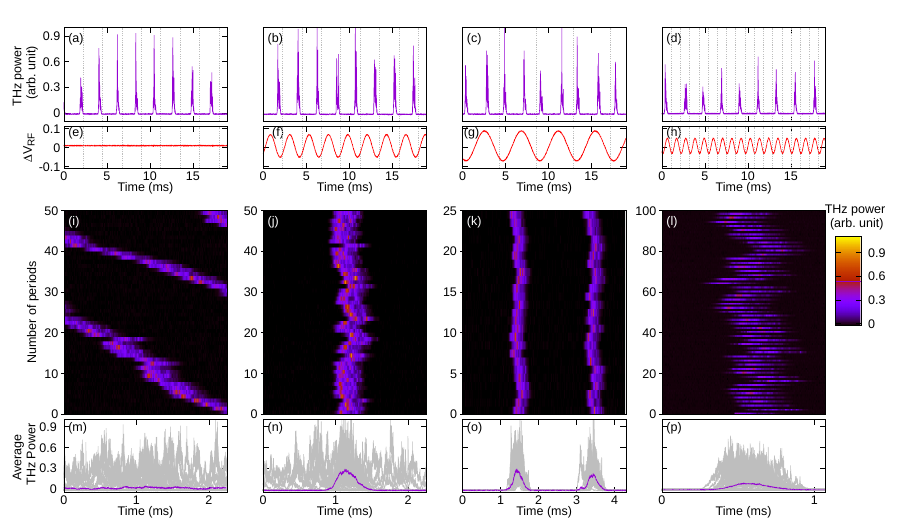}
  \caption{Synchronization phenomena from experimental signals. a-d) Time evolution of the THz power (recorded with a bolometer), when (a) there is no  modulation ($A_{RF}=0$) and the burst period $T_b$ is about of $2.19$~ms (the period used for constructing the map is the same as in the next column, 2.25 ms)  ; (b) the RF modulation period $T_{RF}$ is near the spontaneous burst period $T_b$ ($T_{RF}=2.25~$ms, $A_{RF}=0.03$); (c) $T_{RF}$ is near the half of $T_b$ ($T_{RF}=4.29~$ms, $A_{RF}=0.04$); and (d) $T_{RF}$ is near twice $T_b$ ($T_{RF}=1.07~$ms, $A_{RF}=0.02$). (e-h) associated RF modulation signal $\Delta V_{RF}$. In these figures, vertical grey lines indicates the RF signal period $T_{RF}$. i-l) THz power in the map representation [as in the FIG.~2(d)]. m-p) average value of the THz power (in the map representation) over the number of periods (in the last case, the average is taken every two periods of the RF signal, when there is a burst). Grey lines are the superposition of all the bursts. In all theses plots, a transient of 20~ms (after the application of the RF modulation) is not shown.}

\end{figure*}

We also observe that the modulation signal doesn't change significatively the electron bunch distribution [see  FIG.~1(j), and the entire dynamics in the movie in supplementary material].
This can be understood by noting that this method is only weakly perturbative, since the modulation amplitude is quite small (5\% of the total RF signal amplitude $V_{RF}$).
However, we can observe that the burst shapes are slightly modified due to the modulation [see FIG.~2(b)]. In particular, in this case, the burst have a lower amplitude (due to a higher repetition rate  imposed by the RF signal).\\

Finaly, even if the bursts are synchronized, it doesn't imply  that  the micro-structures between different bursts are in phase (even if these micro-structures are at the origin of the bursts).
This can be observed looking at a modulation in the bursts [see FIG.~2(e) and the inset].
It is known that this modulation is linked to the time apparition of the micro-structures in the longitudinal phase-space~\cite{Roussel2014, Steinmann2018}.
Since the modulation of two consecutive bursts are not superposed, it indicates that the RF signal modulation doesn't synchronize the apparition of the micro-structures.\\

%%%%%%%%% EXP %%%%%%%%%%%%%%%
\section{Experiment}

%%%%% setup %%%%%%%%
The experiment was conducted at the Synchrotron SOLEIL during a dedicated experimental run.
The THz power emitted by the electron bunch was recorded at the THz-IR AILES beamline, using a bolometer (with $1~\mu$s response time).
The storage ring was operated in a single-bunch mode, with the circulating bunch carrying a current above instability threshold (of about 12~mA, and thanks to the top-up injection scheme, the current remained quasi-constant throughout the measurement). The instability threshold is of about 11.3~mA.
Under these conditions, the bunch emits coherent synchrotron radiation (CSR) in the THz range, exhibiting a bursting behavior, as shown in Fig.~3(a).
The characteristic period of these spontaneous bursts $T_b$ is approximately 2.19~ms.\\

\begin{figure*}[htpb!]
  \includegraphics[width=0.9\linewidth]{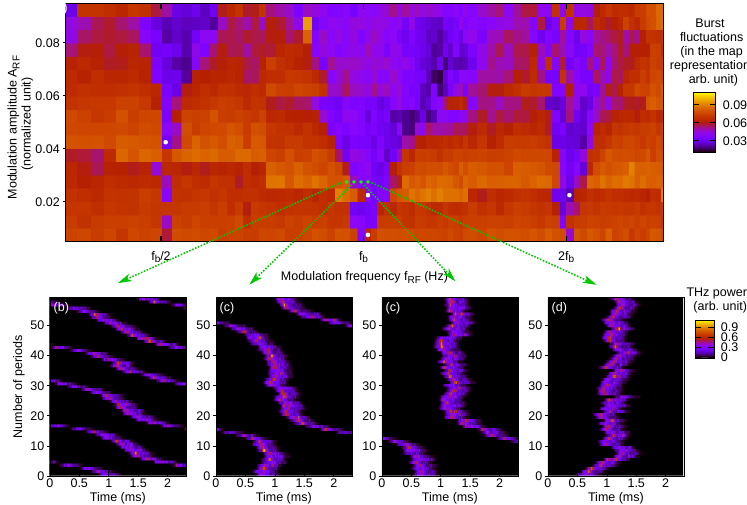}
  \caption{a) Fluctuations of the THz power $\Delta P$ in function of the frequency modulation $f_{RF}$ (in logscale), and in function of the modulation amplitude $\Delta V_{RF}$,  in the map reprensentation (with a transient of xx~ms removed). With this representation, three Arnold tongues appears (for the sub-harmonic case centered around $f_b/2$, for the fondamental case centered around $f_b$ and for the harmonic case, centered around $2 f_b$, with $f_b$ the spontaneous burst frequency  (i.e. without modulation). ). Positions of the four cases of the FIG.~3 are indicated by the white circle.
    b-e) THz power in the map representation, for four different frequency modulations ($f_{RF}=432.9$~Hz,$444~$Hz, $455~$Hz, $466$~Hz) crossing the border of the fundamental arnold tongue and for a fixed amplitude modulation $\Delta V_{RF}=0.025$. Positions of these four cases are indicated in green circles in the figure~a). In these figure, the transient is displayed (the modulation is activated at $t=0$ of the first period).  todo: check burst frequency (fb) for deltaV=0.}   \label{fig:dyn_sync}
\end{figure*}

The modulation signal was generated by an FPGA card (RedPitaya STEMlab 125-14 board model), with a homemade code employing the DDS (Direct Digital Synthesis) method, with parameters controlled from an external PC.
This modulation signal was injected into the low-level RF system of the RF cavity.
The cavity was tuned in the so-called "zero-crossing" configuration  [see Fig.~1(d)], which allows variations in the amplitude of the RF signal to be directly translated into variations in the slope of the electric field experienced by the electron bunch as it traverses the cavity (and thus it permits to modify the bunch length). Three other cavities operate in standard mode to accelerate the electrons [see FIG.~1(c)], each with an amplitude of 0.5~MV (and the cavity operated in zero-crossing mode also has an amplitude of 0.5 MV when $A_{RF} = 0$).\\

The figure~3 shows the effects of this modulation for several cases.
First case [see FIG.~3(a),(e),(i), (m)] corresponds to a case without synchronization, thus associated to a drift of the burst position in the map representation.
The spontenous period of the bursts $T_b$ can be measured and is about of $2.19~$ms. 
The next case [see FIG.~3(b),(f),(j), (n)] corresponds to a synchronization at a frequency close to the spontaneous burst frequency (with $T_{RF}=2.25$~ms and a normalized amplitude of $A_{RF}=0.03$, corresponding to $60~$kV).
Then, as is characteristic of such synchronization processes, it is also possible to achieve synchronization at harmonics or subharmonics of the modulation frequency.
The case of the first harmonic is shown in Fig.~3(c),(g),(k), (o)] (with $T_{RF}=4.29~$ms, $A_{RF}=0.04$), where two CSR bursts occur during one modulation period.
The case of the first subharmonic is shown in Fig.~3(d),(h),(l),(p) (with $T_{RF}=1.07~$ms, $A_{RF}=0.02$), where one CSR burst occurs every two modulation periods.
In these three cases, even if a synchronization is clearly acheived, the burst amplitudes still fluctuate [see FIG.~3(b)-(d)].\\

%%%%%% Arnold Tongues %%%%

To evaluate the efficiency of synchronization as a function of the RF signal modulation parameters — the amplitude $A_{RF}$ and the period $T_{RF}$ — systematic recordings were performed and subsequently analyzed by calculating the fluctuations of the THz power using the map representation.
In this framework, the THz power associated with the $i$-th period is denoted by $P_i(t')$, where $t'$ represents the time within a given period (ranging from $0$ to $T_{RF}$).
The fluctuations $\Delta P$ are then quantified using the following expression (RMS fluctuations):

\begin{equation}
  \Delta P  = \frac{1}{N} \sqrt{  \int_0^{T_{RF}}  \sum_{i=1}^N \left[ \left\langle P(t') \right\rangle - P_i(t') \right]^2 dt'},
\end{equation}

where $N$ is the total number of periods and $\left\langle P(t') \right\rangle$ is the average THz power at a given time $t'$ over all periods~$i$ $\left( \left\langle P(t') \right\rangle = \frac{1}{N} \sum_{i=1}^N P_i(t') \right)$.
This average power  is shown for the different cases of the figure~3 [in the FIG.~3~(m–p)]. Note that for the part associated to the harmonic case, the period used to calculate $\Delta P$  is $2\times T_{RF}$, since there is one burst every two periods [as it can be seen in FIG.~3(l)].\\

The calculated fluctuations provide a quantitative criterion to distinguish between synchronized and unsynchronized regimes: low fluctuations correspond to synchronized states, where the THz bursts are aligned temporally in the map representation, whereas high fluctuations indicate a lack of synchronization, with bursts appearing at varying times in the map representation.
These fluctuations are computed as a function of the RF modulation parameters ($A_{RF}$ and $T_{RF}$), and the results are presented in Figure~4.
We observe that the fluctuation map exhibits structures characteristic of Arnold tongues~\cite{Arnold1961}, i.e., synchronization domains that broaden with increasing modulation amplitude $A_{RF}$. In this figure, three Arnold tongues are present, associated to the fundamental (central tongue), sub-harmonic (left tongue) and harmonic case (right tongue ). 
It is worth noting that, even at its maximum, the applied modulation amplitude remains relatively low in comparison to values used in similar studies (e.g., $A_{RF} = 0.8$ in~\cite{Evain2024}).\\

%%%%%% transient duration &  Decrochage/raccorchage %%%%%

Finally, we investigate the different dynamical behaviors that occur when crossing the boundary between synchronized and unsynchronized regimes, for a fixed modulation amplitude of $A_{RF} = 0.025$.
Several regimes are observed for different values of the frequency detuning $\Delta f = f_{RF} - f_b$, where $f_{RF}$ is the RF modulation frequency and $f_b$ is the spontaneous bursting frequency [see FIG.~4(b-e)].
When $\Delta f$ is small (see Fig.~4(e), synchronization is achieved with a short transient.
For a slightly larger detuning (see Fig.~4(d), synchronization still occurs but is preceded by a longer transient regime.
As $\Delta f$ increases further (Fig.~4(c), we observe phase-slip phenomena, a well-known behavior in synchronization theory~\cite{Pikovsky2003}, where the system alternante betwen period of temporarily synchronization, and no synchronuzation. 
For even larger values of $\Delta f$, no synchronization is observed, as shown in Fig.~4(b).\\

%%%%%%%%%%% conclusion ########
\section{Conclusion}
In conclusion,  a synchronization phenomenon occurring in an accelerator-based light source is presented.
This is obtained in a storage ring, where a systematic spatio-temporal instability appears when high bunch charges are used, characterized by the spontaneous appearance of micro-structures within an electron bunch.
These structures are induced by the interaction between the electrons and their radiation, leading to the emission of intense coherent radiation in the terahertz range. \\

These micro-structures appear with a periodic or pseudo-periodic behavior (the so-called bursting behavior), and in this study, we show that it is possible to synchrotronize them with an external sinusoidal signal that modulate a RF cavity field.
Typical synchronization phenomena are observed, such as the appearance of Arnold tongues at the fundamental, harmonic, and subharmonic frequencies of the natural bursting frequency.
A phase-slip behavior is also observed when the system is near the synchronization regime.\\

This study reveals that a relatively small RF modulation amplitude can have a significant impact on the bursting dynamics.
While the present analysis has been conducted in single-bunch mode operation, the approach is, in principle, extendable to multi-bunch operation, allowing for the synchronization of bursts emitted by different electron bunches using a common external RF signal modulation.

\section{Acknowledgement}

The PhLAM team was supported by the following funds: Ministry of Higher Education and Research, Nord-Pas de Calais Regional Council and European Regional Development Fund (ERDF)
through the Contrat de Plan \'Etat-R\'egion (CPER Wavetech).
The authors acknowledge the support of the CDP C2EMPI, together with the French State under the France2030 programme, the University of Lille, the Initiative of Excellence of the University of Lille, the European Metropolis of Lille for their funding and support of the R-CDP-24-004-C2EMPI project.
%Twac

% ANR-DFG ULTRASYNC project (ANR-19-CE30-0031).
\bibliographystyle{plain}

%\bibliography{1}

\begin{thebibliography}{10}

\bibitem{Arnold1961}
V.~I. Arnol'd.
\newblock Small denominators. i. mapping the circle onto itself.
\newblock {\em Izv. Akad. Nauk SSSR Ser. Mat.}, 25(1):21--86, 1961.

\bibitem{Balanov2009}
A.~Balanov, N.~Janson, D.~Postnov, and Olga Sosnovtseva.
\newblock {\em Synchronization: From Simple to Complex}.
\newblock Springer, 2009.

\bibitem{Byrd2006}
J.~M. Byrd, Z.~Hao, M.~C. Martin, D.~S. Robin, F.~Sannibale, R.~W. Schoenlein,
  A.~A. Zholents, and M.~S. Zolotorev.
\newblock Tailored terahertz pulses from a laser-modulated electron beam.
\newblock {\em Phys. Rev. Lett.}, 96:164801, Apr 2006.

\bibitem{Evain2023}
C.~Evain, F.~Kaoudoune, E.~Roussel, C.~Szwaj, M.-A. Tordeux, F.~Ribeiro,
  M.~Labat, N.~Hubert, J.-B. Brubach, P.~Roy, and S.~Bielawski.
\newblock Stabilization of the bunch position during the control of the
  microbunching instability in storage rings.
\newblock {\em Phys. Rev. Accel. Beams}, 26:090701, Sep 2023.

\bibitem{Evain2024}
C.~Evain, E.~Roussel, S.~Bielawski, M.-A. Tordeux, F.~Ribeiro, M.~Labat,
  N.~Hubert, J.-B. Brubach, P.~Roy, and C.~Szwaj.
\newblock Gain switching of the microbunching instability to produce giant
  bursts of terahertz coherent synchrotron radiation.
\newblock {\em Phys. Rev. Lett.}, 133:145001, Oct 2024.

\bibitem{Evain2017}
C~Evain, E~Roussel, M~{Le Parquier}, C~Szwaj, M-A Tordeux, J-B Brubach,
  L~Manceron, P~Roy, and S~Bielawski.
\newblock Direct observation of spatiotemporal dynamics of short electron
  bunches in storage rings.
\newblock {\em Physical Review Letters}, 118(5):054801, 2017.

\bibitem{Evain2019}
C.~Evain and {C. and Roussel, E. et al.} Szwaj.
\newblock Stable coherent terahertz synchrotron radiation from controlled
  relativistic electron bunches.
\newblock {\em Nature Physics}, 15:635, 2019.

\bibitem{Glass1983}
Leon Glass, Michael~R. Guevara, Alvin Shrier, and Rafael Perez.
\newblock Bifurcation and chaos in a periodically stimulated cardiac
  oscillator.
\newblock {\em Physica D: Nonlinear Phenomena}, 7(1):89--101, 1983.

\bibitem{Golden2021}
Alexander Golden, Allyson~E. Sgro, and Pankaj Mehta.
\newblock Arnold tongues in oscillator systems with nonuniform spatial driving.
\newblock {\em Phys. Rev. E}, 103:042211, Apr 2021.

\bibitem{Pikovsky2003}
A~Pikovsky, M.~Rosenblum, and J.~Kurths.
\newblock {\em Synchronization: A Universal Concept in Nonlinear Sciences}.
\newblock Cambridge University Press, 2003.

\bibitem{Pikovsky2001}
Arkady Pikovsky, Michael Rosenblum, and Jürgen Kurths.
\newblock {\em Synchronization: A Universal Concept in Nonlinear Sciences}.
\newblock Cambridge Nonlinear Science Series. Cambridge University Press, 2001.

\bibitem{Pikovsky1997}
Arkady~S. Pikovsky, Michael~G. Rosenblum, Grigory~V. Osipov, and Jürgen
  Kurths.
\newblock Phase synchronization of chaotic oscillators by external driving.
\newblock {\em Physica D: Nonlinear Phenomena}, 104(3):219--238, 1997.

\bibitem{Roussel2015}
E.~Roussel, C.~Evain, M.~Le~Parquier, C.~Szwaj, S.~Bielawski, L.~Manceron,
  J.-B. Brubach, M.-A. Tordeux, J.-P. Ricaud, L.~Cassinari, M.~Labat, M.-E.
  Couprie, and P.~Roy.
\newblock Observing microscopic structures of a relativistic object using a
  time-stretch strategy.
\newblock {\em Scientific Reports}, 5(1):10330, May 2015.

\bibitem{Roussel2014b}
E~Roussel, C~Evain, M~Le Parquier, C~Szwaj, S~Bielawski, M~Hosaka, N~Yamamoto,
  Y~Takashima, M~Shimada, M~Adachi, H~Zen, S~Kimura, and M~Katoh.
\newblock Transient response of relativistic electron bunches to wave-number
  selected perturbations near the micro-bunching instability threshold.
\newblock {\em New Journal of Physics}, 16(6):063027, jun 2014.

\bibitem{Roussel2014}
E.~Roussel, C.~Evain, C.~Szwaj, and S.~Bielawski.
\newblock Microbunching instability in storage rings: Link between phase-space
  structure and terahertz coherent synchrotron radiation radio-frequency
  spectra.
\newblock {\em Phys. Rev. ST Accel. Beams}, 17(010701), 2014.

\bibitem{Ruhunusiri2012}
W.~D.~Suranga Ruhunusiri and J.~Goree.
\newblock Synchronization mechanism and arnold tongues for dust density waves.
\newblock {\em Phys. Rev. E}, 85:046401, Apr 2012.

\bibitem{Schoenfeldt2017}
Patrik Sch\"onfeldt, Miriam Brosi, Markus Schwarz, Johannes~L. Steinmann, and
  Anke-Susanne M\"uller.
\newblock Parallelized vlasov-fokker-planck solver for desktop personal
  computers.
\newblock {\em Phys. Rev. Accel. Beams}, 20:030704, Mar 2017.

\bibitem{Steinmann2018}
Johannes~L. Steinmann, Tobias Boltz, Miriam Brosi, Erik Br\"undermann, Michele
  Caselle, Benjamin Kehrer, Lorenzo Rota, Patrik Sch\"onfeldt, Marcel Schuh,
  Michael Siegel, Marc Weber, and Anke-Susanne M\"uller.
\newblock Continuous bunch-by-bunch spectroscopic investigation of the
  microbunching instability.
\newblock {\em Phys. Rev. Accel. Beams}, 21:110705, Nov 2018.

\bibitem{Venturini2002}
M.~Venturini and R.~Warnock.
\newblock Bursts of coherent synchrotron radiation in electron storage rings: A
  dynamical model.
\newblock {\em Phys. Rev. Lett.}, 89(22):224802, Nov 2002.

\bibitem{Venturini2005}
M.~Venturini, R.~Warnock, R.~Ruth, and J.~A. Ellison.
\newblock Coherent synchrotron radiation and bunch stability in a compact
  storage ring.
\newblock {\em Phys. Rev. ST Accel. Beams}, 8(1):014202, Jan 2005.

\bibitem{Warnock:2006qa}
Robert~L. Warnock.
\newblock {Study of bunch instabilities by the nonlinear Vlasov-Fokker-Planck
  equation}.
\newblock {\em Nucl. Instrum. Meth. Phys. Res., Sect. A}, 561:186--194, 2006.

\end{thebibliography}

\section{Appendix: Numerical model}
Numerical simulations are carried out with the same model and numerical implementation as in~[evain2019], except for the introduction of the RF–signal modulation.  
The adopted framework is the Vlasov--Fokker--Planck model for storage rings~\cite{Warnock:2006qa}, which is known to accurately reproduce the microbunching instability~\cite{Schoenfeldt2017,Roussel2015}.  
When including a modulation of the RF amplitude, the governing equation becomes:

\begin{eqnarray} \textstyle
  \frac{\partial f(q,p,\theta)}{\partial \theta} &-& \textstyle  p \frac{\partial f}{\partial q} +  \frac{\partial f}{\partial p} \left[ q (1 + \Delta V_{RF}(\theta))   - I_c E^e_{\mathrm{THz}} \right]\\
  &=& 2\epsilon \frac{\partial}{\partial p}\left( p f + \frac{\partial f }{\partial p}\right)
 \end{eqnarray}

Here, $q$ denotes the normalized longitudinal position ($q = z/\sigma_z$, with $z$ the longitudinal coordinate and $\sigma_z$ the equilibrium RMS bunch length in the absence of collective effects), and $p$ is the normalized longitudinal energy ($p = (E - E_0)/\sigma_E$, where $E$ is the electron energy, $E_0$ the nominal storage–ring energy, and $\sigma_E$ the equilibrium energy spread without collective effects).  
The variable $\theta$ corresponds to the normalized time $\theta = t/(2\pi f_s)$, where $t$ is the physical time and $f_s$ the synchrotron frequency.  
The parameter $\epsilon$ is defined as $\epsilon = 1/(2\pi f_s \tau_s)$, with $\tau_s$ the synchrotron damping time.\\

The collective interaction term responsible for the instability, $E^{B}_{\mathrm{THz}}$, is computed using the standard parallel–plate impedance model (plate radius $R_c$ and separation $2h$), as detailed in~\cite{Evain2017}.  
The normalized bunch current is defined as $I_c = I \frac{e \, 2\pi R_c}{2\pi f_s \sigma_E T_0}$, where $I$ is the average beam current, $R_c$ the dipole bending radius, $T_0$ the revolution period, and $e$ the electron charge. All quantities are expressed in MKS units.\\

The RF amplitude modulation is taken to be sinusoidal:
\[
\Delta V_{RF}(\theta) = A_{RF} \sin\!\left( \frac{2\pi}{T'_{RF}} \theta \right),
\]
where $T'_{RF}$ denotes the modulation period in normalized units, 
$T'_{RF} = T_{RF}/(2\pi f_s)$, with $T_{RF}$ the physical modulation period.\\

The emitted THz power $P_{\mathrm{THz}}$ is computed using the standard expression: 
\[
P_{\mathrm{THz}}(\theta) = \int_{k_0}^{\infty} |\tilde{\rho}(k,\theta)|^2 \, dk,
\]
where $\tilde{\rho}(k)$ is the Fourier transform of the charge density $\rho(q) = \int_{-\infty}^{\infty} f(q,p)\, dp$, and $k_0$ is the cutoff wavenumber associated with the bolometer detection bandwidth.

The parameters used in the article (SOLEIL synchrotron) are:
$E_0=2.75~$GeV, $\sigma_z=4.59~$mm, $\sigma_E=1.0168\times 10^{-3} E_0$, $f_s=4.64~$kHz, $\tau_s=3.27~$ms, $R_c=5.36~$m, $T_0=1.181~\mu$s, $h=1.25~$cm, $I=10~$mA,  $k_0=2.5$.

\end{document}